\documentclass{aa}  
\usepackage{graphicx}
\usepackage{txfonts}
\usepackage{hyperref}

\usepackage{xcolor}
\usepackage[normalem]{ulem}
\usepackage{orcidlink}

\defcitealias{2025A&A...701A.169S}{Paper\,I}

\begin{document} 

\title{JWST imaging of omega\,Centauri}
\subtitle{II. Evidence for a split white dwarf cooling sequence in the near-infrared}

\author{M.\,Scalco\inst{1}\orcidlink{0000-0001-8834-3734}\fnmsep\thanks{\email{mscalco@iu.edu}}
\and 
M.\,Salaris\inst{2,3}\orcidlink{0000-0002-2744-1928}
\and
L.\,R.\,Bedin\inst{4}\orcidlink{0000-0003-4080-6466}
\and
S.\,Blouin\inst{5}\orcidlink{0000-0002-9632-1436}
\and
E.\,Vesperini\inst{1}
\and
P.\,Bergeron\inst{6}\orcidlink{0000-0003-2368-345X}
\and
M.\,Libralato\inst{4}\orcidlink{0000-0001-9673-7397}
\and
M.\,Griggio\inst{7}\orcidlink{0000-0002-5060-1379}
\and
A.\,Burgasser\inst{8}\orcidlink{0000-0002-6523-9536}
\and
D.\,Nardiello\inst{9,4}\orcidlink{0000-0003-1149-3659}
\and 
A.\,Bellini\inst{7}
\and
J.\,Anderson\inst{7}
\and
R.\,Gerasimov\inst{10}\orcidlink{0000-0003-0398-639X}
\and
D.\,Apai\inst{11,12}\orcidlink{0000-0003-3714-5855}
}

\institute{
Department of Astronomy, Indiana University, Swain West, 727 E. 3rd Street, Bloomington, IN 47405, USA
\and
Astrophysics Research Institute, Liverpool John Moores University, 146 Brownlow Hill, Liverpool L3 5RF, UK
\and
Istituto Nazionale di Astrofisica, Osservatorio Astronomico di Abruzzo, via M. Maggini sn, I-64100 Teramo, Italy
\and
Istituto Nazionale di Astrofisica, Osservatorio Astronomico di Padova, Vicolo dell'Osservatorio 5, Padova I-35122, Italy
\and
Department of Physics and Astronomy, University of Victoria, Victoria, BC V8W 2Y2, Canada
\and
D{\'e}partement de Physique, Universit{\'e} de Montr{\'e}al, C.P. 6128, Succ. Centre-Ville, Montr{\'e}al, Quebec H3C 3J7, Canada
\and
Space Telescope Science Institute, 3700 San Martin Drive, Baltimore, MD 21218, USA
\and
Department of Astronomy \& Astrophysics, University of California, San Diego, La Jolla, California 92093, USA
\and
Dipartimento di Fisica e Astronomia "Galileo Galilei", Universit{\`a} di Padova, Vicolo dell'Osservatorio 3, Padova I-35122, Italy
\and
Department of Physics and Astronomy, University of Notre Dame, Nieuwland Science Hall, Notre Dame, Indiana 46556, USA
\and
Department of Astronomy and Steward Observatory, The University of Arizona, 933 N. Cherry Avenue, Tucson, AZ 85721, USA
\and
Lunar and Planetary Laboratory, The University of Arizona, 1629 E. University Blvd., Tucson, AZ 85721, USA
}

\date{XXX,YYY,ZZZ}
 
\abstract
{We present a detailed analysis of the white dwarf cooling sequence (WD CS) in omega\,Centauri based on combined \textit{Hubble} Space Telescope (HST) and JWST observations. Our analysis confirms the previously reported split --based on HST observations in ultraviolet filters-- in the upper part of the WD CS, consistent with the presence of two distinct WD populations, and extends it to a significantly fainter and cooler limit (down to $\sim8{,}000$\,K), corresponding to cooling ages of about 1\,Gyr. We used artificial star (AS) tests and cooling models to confirm that the split is evidence of two WD populations with different masses and progenitors: one sequence of \lq{canonical\rq} WDs produced by the He-normal progenitors, and one sequence of low-mass WDs originated from the cluster He-rich component. We show that the fraction of WDs from the He-rich component in the outer regions is smaller than that found in the innermost regions. We also studied the kinematics of WDs and showed that in the outer regions, the velocity distribution of WDs from He-rich progenitors is slightly radially anisotropic, while that of canonical WDs is slightly tangentially anisotropic. Both the radial variation of the fraction of WDs from the He-rich population and the difference between their velocity distribution and that of canonical WDs are consistent with spatial and kinematic differences previously found for He-rich and He-normal main-sequence (MS) stars and in general agreement with models predicting that He-rich stars form more centrally concentrated than He-normal stars. 
}

\keywords{Proper motions, globular clusters: individual: NGC\,5139, white dwarfs}

\titlerunning{JWST imaging of the split WD sequence in $\omega$\,Cen}
\authorrunning{M.\,Scalco et al.}
\maketitle

\section{Introduction}\label{Section1}

White dwarfs (WDs) are the evolutionary endpoints of low- and intermediate-mass stars, and their cooling sequences (CSs) provide powerful diagnostics for stellar populations, particularly in globular clusters (GCs; e.g. \citealt{1952MNRAS.112..583M,1987ApJ...315L..77W,2000ApJ...529..318R,2001PASP..113..409F,2009ApJ...692.1013S}). The WD CS not only serves as an age indicator through its faint end but also retains signatures of the evolutionary history and chemical composition of the progenitor populations, making it a particularly valuable tool in the study of massive and dynamically evolved GCs \citep{2007ApJ...671..380H,2009ApJ...697..965B,2019MNRAS.488.3857B,2023MNRAS.518.3722B,2024AN....34540039B,2025AN....34640125B,2013ApJ...778..104R}. 

The massive GC Omega\,Centauri (NGC\,5139, hereafter $\omega$\,Cen) has long been recognised as an exceptionally complex system \citep{1966ROAn....2....1W}, hosting multiple distinct stellar populations. These populations show up across all evolutionary phases, from the main sequence (MS) to the red giant branch (RGB), and into the WD CS \citep[e.g.][and reference therein]{1999Natur.402...55L,2000ApJ...534L..83P,2004ApJ...605L.125B,2024A&A...691A..96S}.

The upper part of the WD CS down to $T_{\rm eff} \simeq 15{,}500$\,K, which includes the brightest, and least massive recently formed WDs in $\omega$\,Cen, was first reported to split into two distinct branches by \citet{2013ApJ...769L..32B}, based on ultraviolet observations with the \textit{Hubble} Space Telescope (HST) in a central field of the cluster. These two branches were identified as a blue CS, primarily composed of standard $0.53-0.55\,{\rm M}_{\odot}$ CO-core WDs, and a red CS, interpreted as a mixture of lower-mass CO-core WDs and He-core WDs.

The presence of a significant population of He-core WDs had already been identified by  \citet{2005ApJ...621L.117M} and \citet{2007ApJ...663.1021C}. These differences in WD properties reflect the varying evolutionary histories of their progenitors: the red WD CS likely descends from He-enriched subpopulations that produced the extremely blue horizontal branch (HB) stars observed in the cluster, while the blue WD CS traces canonical He-normal populations. 

Previous studies of $\omega$\,Cen have found differences between the dynamical properties of He-rich and He-normal MS stars, providing insight into the formation and evolutionary history of the multiple populations of this cluster. The study of the spatial distribution and kinematic properties of the two populations of WDs and the comparison with those of MS stars may offer additional insight into the cluster formation and evolution, and provide dynamical evidence of the link between the WD populations and the He-rich and He-normal MS stars.

This work extends previous studies by tracing the WD sequences in $\omega$\,Cen to fainter magnitudes and cooler temperatures than previously explored. Our main focus is to determine down to which magnitude the split between the two WD sequences remains clearly detectable, providing insights into the formation epoch of the extreme-HB progenitors. In addition, we investigate how the relative fraction of the two WD populations varies with radial distance from the cluster centre, study their kinematics, and compare these behaviours with the known radial distribution of the MS subpopulations.

Previous investigations of the WD CS in $\omega$\,Cen have been conducted primarily at ultraviolet and optical wavelengths. The unprecedented sensitivity and spatial resolution of JWST now allow us to investigate this sequence also in the near-infrared (NIR), opening a new observational window onto the faintest WDs in the cluster.

This study is part of a broader series aimed at mapping the imprint of multiple stellar populations (mPOPs) in $\omega$\,Cen using JWST. In \citet[][hereafter \citetalias{2025A&A...701A.169S}]{2025A&A...701A.169S}, we investigated the luminosity and mass functions of the cluster's two main stellar populations along the MS. Here in Paper\,II, we focus on the WD CS, using a combination of JWST and HST data to analyse the split between the two WD populations, their radial distribution, kinematics, and the connection with their progenitor MS populations.

The paper is organised as follows: Section\,\ref{Section2} describes the data and reduction methods. Section\,\ref{Section3} presents the artificial star (AS) tests used to estimate the photometric uncertainties. Section\,\ref{Section4} analyses the WD CS morphology, its intrinsic broadening, and population decomposition, including comparison with theoretical isochrones, radial gradients and kinematics. A summary of the main results is provided in Section\,\ref{Section6}.

\section{Data set and reduction}\label{Section2}
The data used in this study were presented in \citetalias{2025A&A...701A.169S}, to which we refer for a detailed description of the dataset and reduction process.

Briefly, the dataset combines observations from HST and JWST. The HST images come from the primary field of the GO-14118+14662 multi-epoch programme \citep{2016hst..prop14118B,2016hst..prop14662B} and were acquired with the Wide Field Channel (WFC) of the Advanced Camera for Surveys (ACS) using the F606W and F814W filters, spanning a period from $\sim2015.6$ to $\sim2018.5$. We adopted for these HST data the high-quality catalogue presented in \citet{2024A&A...691A..96S}, which provided a dedicated study of the WD CS in $\omega$\,Cen. This catalogue, specifically optimised for the analysis of the WD CS, provided the first complete coverage of the sequence in $\omega$\,Cen, down to the peak at its termination. A detailed description of the dataset and the reduction process is available in \citet{2024A&A...691A..96S}.

The JWST data originate from the GO-5110 programme \citep[][]{2024jwst.prop.5110B} and were obtained with the Near Infrared Camera (NIRCam). Observations were performed simultaneously with the F150W2 filter of the Short Wavelength (SW) channel and the F322W2 filter of the Long Wavelength (LW) channel at epoch $\sim2024.6$.

Data reduction of the images involved a combination of first- and second-pass photometry for both HST \citep[see also][for details]{2017ApJ...842....6B,2018ApJ...853...86B,2018MNRAS.481.3382N,2018ApJ...854...45L,2022ApJ...934..150L,2021MNRAS.505.3549S} and JWST data \citep[see also][for details]{2024AN....34540039B, 2025AN....34640125B, 2024A&A...689A..59S, 2025A&A...694A..68S, 2022MNRAS.517..484N, 2023MNRAS.525.2585N, 2023MNRAS.521L..39N, 2023AN....34430006G, 2023ApJ...950..101L, 2024PASP..136c4502L}. 

The astrometry was anchored to the absolute reference frame provided by \textit{Gaia} Data Release 3 \citep[DR3;][]{2016A&A...595A...1G,2023A&A...674A...1G}, and photometry was calibrated to the Vega-magnitude photometric system following the prescription of \citet{2005MNRAS.357.1038B} for HST and \citet{2023MNRAS.525.2585N} for JWST.

\begin{centering} 
\begin{figure*}[h!]
 \includegraphics[width=\textwidth]{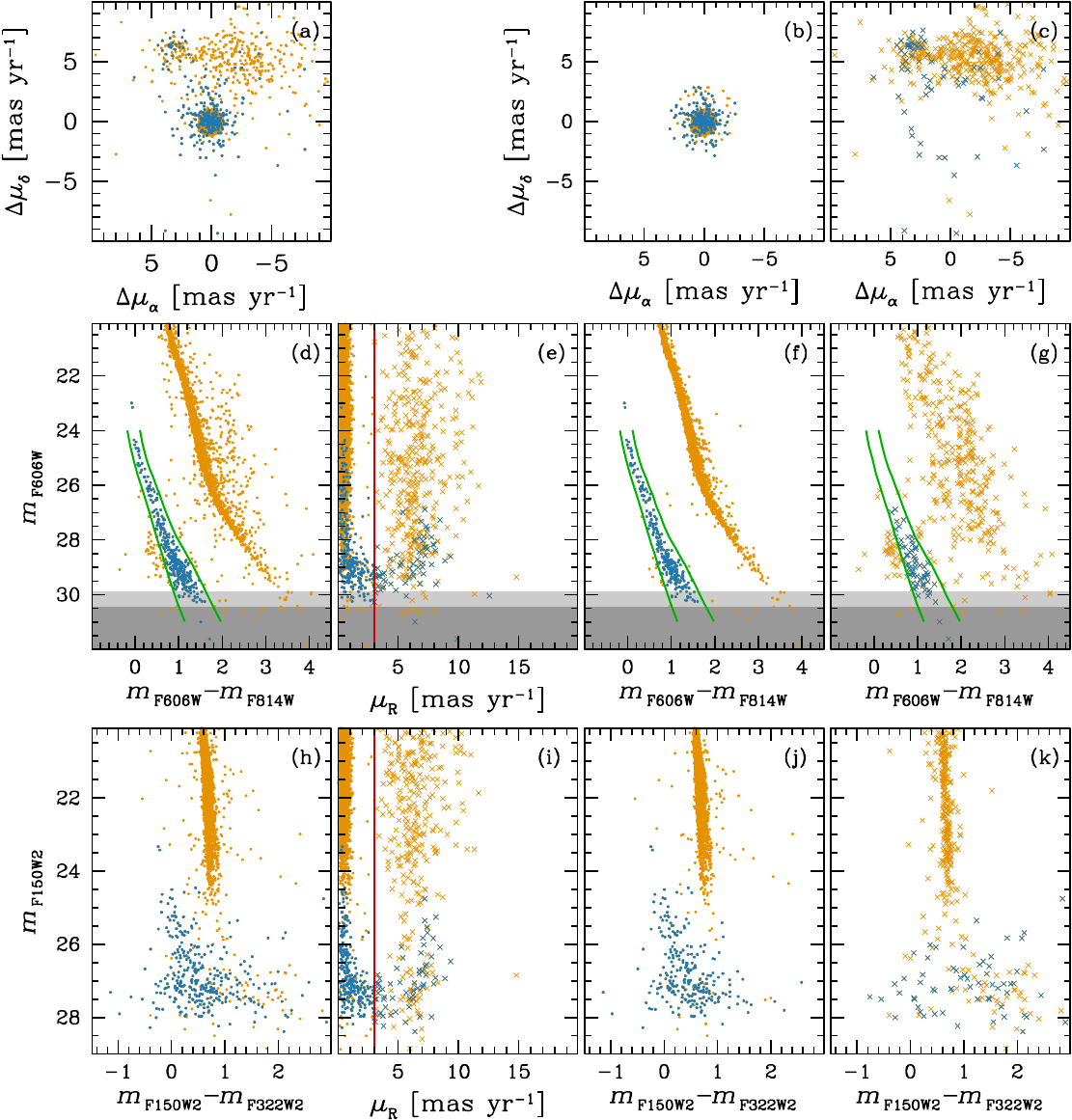}
 \caption{Proper-motion analysis of the sources in the field. Only stars that passed the photometric quality selections and have measurable PMs are included. 
 (a) VPD for all selected sources. (b) VPD for sources that satisfy the PM selection criterion ($\mu_{\rm R} < 3$\,mas\,yr$^{-1}$; see panels\,e and i). (c) VPD for sources that do not meet the PM selection. (d) HST-based CMD for all selected sources, where the WD CS is highlighted in blue within the green fiducial lines. The 5$\sigma$ and 3$\sigma$ detection limits are marked in light and dark grey, respectively. (e) One-dimensional PM ($\mu_{\rm R}$) as a function of the HST magnitude ($m_{\rm F606W}$), with the PM selection threshold (red vertical line) separating cluster members from field stars. (f) Same CMD as (d), but only for stars that pass the PM selection. (g) Same CMD as (d), but only for stars that fail the PM selection. (h) JWST-based CMD for all selected sources. (i) One-dimensional PM ($\mu_{\rm R}$) as a function of the JWST magnitude ($m_{\rm F150W2}$). (j) Same CMD as (h), but only for stars that pass the PM selection. (k) Same CMD as (h), but only for stars that fail the PM selection. In all panels except (a), (d), and (h), cluster members (i.e., stars with $\mu_{\rm R} < 3$\,mas\,yr$^{-1}$) are shown as dots, while non-members are marked with crosses.} 
 \label{pm} 
\end{figure*} 
\end{centering}

We selected a sample of well-measured stars by applying quality-based selection criteria using parameters such as the quality-of-PSF fit \citep[\texttt{QFIT}; see][]{2008AJ....135.2055A}, the excess or deficiency in the source’s flux with respect to the PSF model \citep[\texttt{RADXS}; see][]{2008ApJ...678.1279B}, and the local sky noise \citep[\texttt{rmsSKY}; see][]{2009ApJ...697..965B}. These selection criteria were applied exclusively to the HST data, where the WD CS is more clearly defined. The selection criteria adopted in this work are the same as those defined in \citet{2024A&A...691A..96S}, which were specifically optimised for the study of the WD CS, resulting in the first comprehensive coverage of the sequence down to its peak. In particular, we retained sources with $\texttt{QFIT} > 0$, while for \texttt{RADXS} and \texttt{rmsSKY} we manually defined fiducial thresholds as a function of magnitude, following the trend of each parameter with magnitude, and excluded all sources lying above or below these thresholds. Furthermore, we restricted the sample to stars measured in all four filters.

Proper motions (PMs) were computed using the method described in \citet{2021MNRAS.505.3549S} and introduced in \citet[][see also \citealt{2018ApJ...853...86B,2018ApJ...854...45L,2022ApJ...934..150L}]{2014ApJ...797..115B}. In this iterative approach, each exposure is treated as an independent epoch. The procedure consists of two main steps: first, stellar positions from individual images are transformed into a common reference frame via a six-parameter linear transformation; second, these transformed positions are fitted as a function of time with a least-squares straight line. The slope of the fit, obtained after multiple rounds of outlier rejection, provides the PM measurement. To correct for high-frequency systematic variations, we adopted the method described by \citet{2021MNRAS.505.3549S}, subtracting the median PM of the 100 nearest likely cluster members (excluding the target star itself).

Figure\,\ref{pm} presents the obtained PMs for stars that passed the quality photometric selections and have a measurable PM. Panel\,(a) shows the vector-point diagram (VPD), while panels\,(d) and (h) display the colour-magnitude diagrams (CMDs) based on HST and JWST filters, respectively.

In panel\,(d), and in all panels featuring HST filters, the light grey and dark grey shaded areas denote the 5$\sigma$ and 3$\sigma$ detection limits, respectively, for the sources of interest \citep[see][for details]{2024A&A...691A..96S}. As shown in the CMD of panel\,(d), the WD CS appears to extend down to $m_{\rm F606W} \sim 30.5$, where \citet{2024A&A...691A..96S} identified the termination of the sequence, confirming its actual endpoint. In this CMD, we defined two fiducial lines (shown in green), previously introduced in \citet{2024A&A...691A..96S}, to delineate the WD CS. These fiducial lines were defined by visually inspecting the distribution of real and ASs along the WD CS. They were hand-drawn to strike a balance between encompassing observed WDs with significant photometric scatter and excluding the majority of field objects. In all panels of Fig.\,\ref{pm}, stars within these fiducials are shown in blue, while other stars are displayed in orange.

Panels\,(e) and (i) display the one-dimensional PM ($\mu_{\rm R}$), obtained by combining the PM components in quadrature, plotted against $m_{\rm F606W}$ and $m_{\rm F150W2}$, respectively. These plots show a significant improvement in PM precision compared to \citet[][see Fig.\,3]{2024A&A...691A..96S}, thanks to the longer temporal baseline in this study. This improvement allows for a more robust characterisation of membership along the WD CS. We define as cluster members all sources with $\mu_{\rm R} < 3$, as indicated by the red vertical line in panels\,(e) and (i). Stars that satisfy the PM selection criteria are represented by dots, while those that do not are shown as crosses.

Panels\,(b), (f), and (j) display the same data as panels\,(a), (d), and (h) but only for stars that pass the PM selection, whereas panels\,(c), (g), and (k) show the same information for stars that do not meet the PM selection criteria. It is worth noting that in panel\,(c), the blue crosses appear clustered in a distinct region of the VPD. These sources are most likely faint, blue point-like, very distant galaxies in the background of $\omega$\,Cen. In panel\,(f), we observe a small group of WDs located below the 5$\sigma$ detection limit. These stars are expected and correspond to the peak of the WD CS luminosity function identified in \citet{2024A&A...691A..96S}. Because in this study we restrict the analysis to the sources detected in both the HST and JWST datasets, the number of stars appearing in this faintest region of the sequence is lower than in the previous HST-only work. A few additional sources lying slightly to the left of the bluer fiducial line can also be seen in panel\,(f). These objects are likely either poorly measured cluster WDs or field stars that happened to pass the PM selection.

In what follows, we will only consider the sources shown in panels\,(f) and (j), i.e. those that satisfy both the photometric quality criteria and the PM-based membership selection (panels\,e and\,i).

\begin{centering} 
\begin{figure}
 \includegraphics[width=\columnwidth]{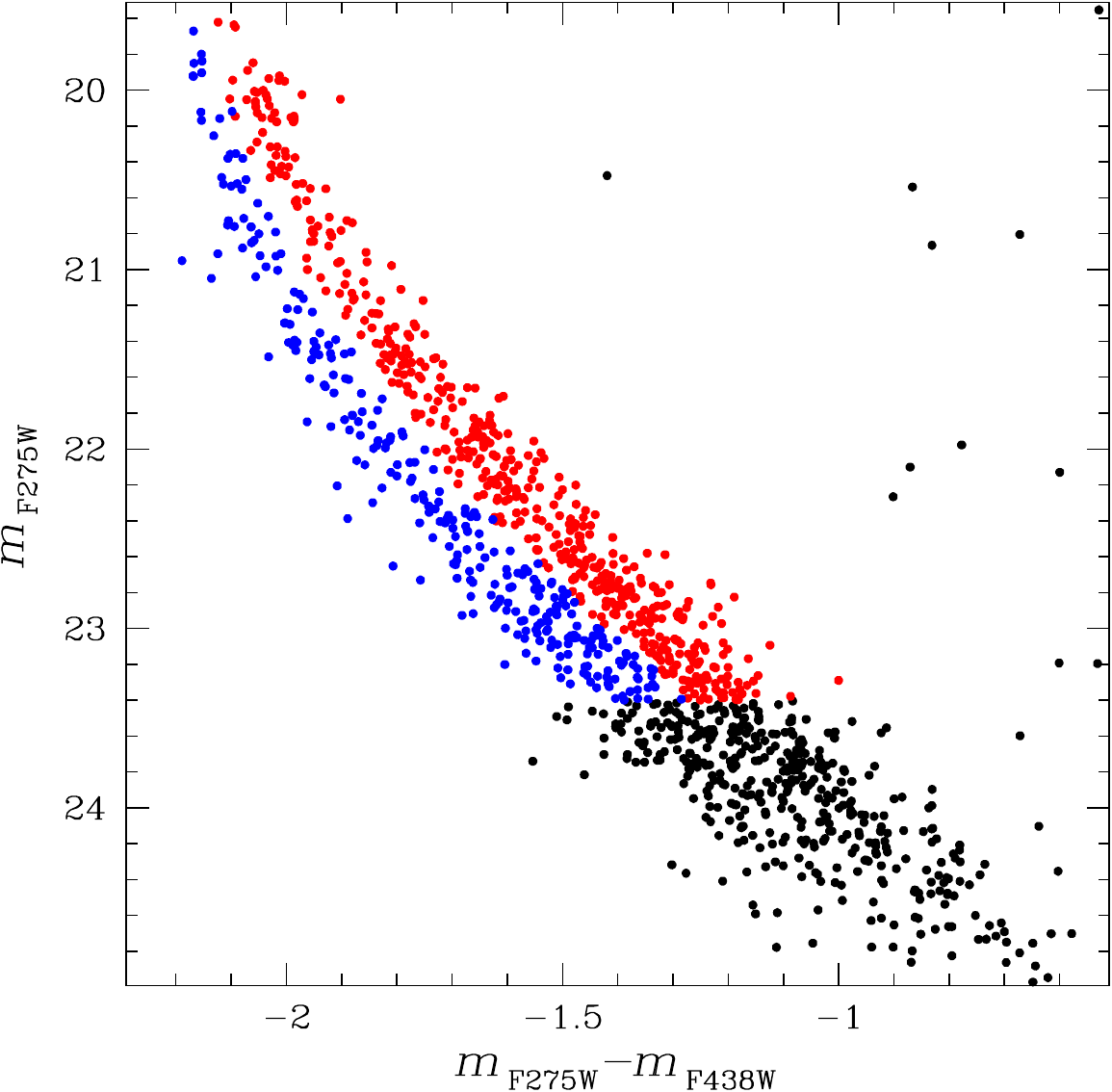}
 \caption{$m_{\rm F275W}$ versus $m_{\rm F275W} - m_{\rm F438W}$ CMD of the central field of $\omega$\,Cen, based on the catalogue from \citet{2014ApJ...797..115B}. These data correspond to the HST observations that enabled the discovery by \citet{2013ApJ...769L..32B} of the split in the upper part of the WD CS, down to an effective temperature of approximately $T_{\rm eff} \sim 15{,}500$\,K.}  
 \label{Bellini1} 
\end{figure} 
\end{centering}

\section{Artificial stars}\label{Section3}

We performed AS tests to estimate the photometric errors in our sample. A total of 10$^5$ ASs were generated, uniformly distributed across the overlapping field of view (FOV) between the JWST and HST datasets. The F150W2 magnitudes of these ASs were uniformly sampled within the range $23 < m_{\rm F150W2} < 30$. The corresponding magnitudes in the F322W2, F606W, and F814W filters were assigned based on fiducial lines manually defined on the $m_{\rm F150W2}$ versus $m_{\rm F150W2}-m_{\rm F322W2}$, $m_{\rm F150W2}$ versus $m_{\rm F606W}-m_{\rm F150W2}$ and $m_{\rm F150W2}$ versus $m_{\rm F814W}-m_{\rm F150W2}$ CMD, respectively. These fiducial lines trace the WD CS, extending to the apparent faint end of the sources, and extrapolated to even fainter magnitudes.

The ASs were generated, detected, and measured following the same procedures applied to the real stars. Artificial PMs were generated by following the same procedure adopted for the real stars\footnote{Spurious positional offsets caused by noise -- such as uncertainties in PSF modelling, coordinate transformations, cosmic ray hits, and detector imperfections -- can affect PM measurements and membership selection, particularly for faint stars. Since ASs are injected with identical positions in both epochs, any measured displacement can be entirely attributed to such noise. This approach allows us to quantify this effect, and include it in the photometric errors evaluation.}.

We followed the methodology described in \citep[][Sect.\,2.3]{2009ApJ...697..965B} to correct for systematic errors between input and output magnitudes in both the real and artificial sources. We found these corrections negligible (<0.1 mag) for the JWST data down to the faintest magnitudes studied; therefore, we applied them only to the HST photometry.

\begin{centering} 
\begin{figure}
 \includegraphics[width=\columnwidth]{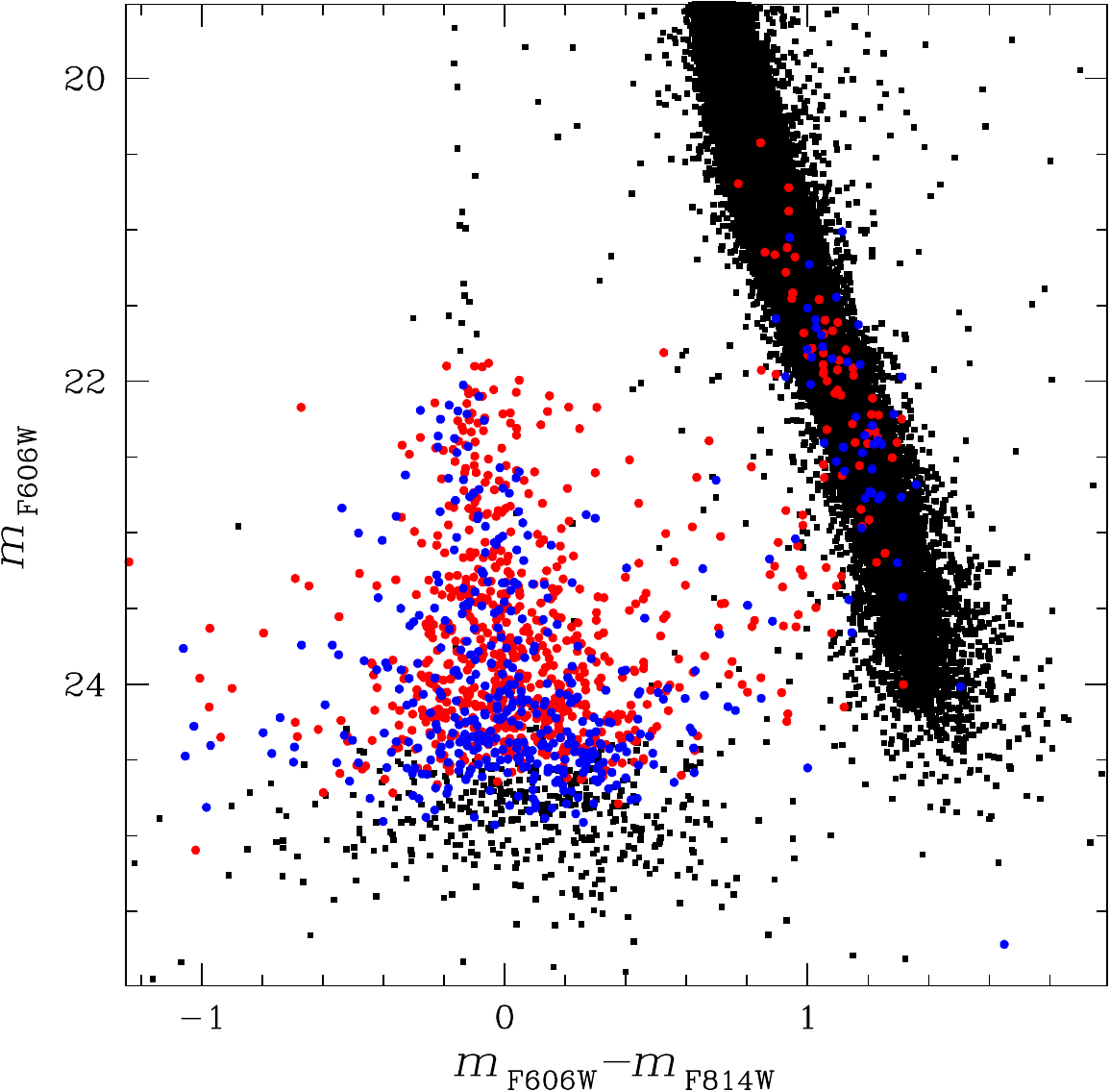}
 \caption{$m_{\rm F606W}$ versus $m_{\rm F606W} - m_{\rm F814W}$ CMD of the stars shown in Fig.\,\ref{Bellini1}, with the same colour-coding. Photometry from \citet{2014ApJ...797..115B}. Note how the two separate groups identified in Fig.\,\ref{Bellini1} appear here as a tangled mixture. Crowding and incompleteness hinder the photometric precision required to resolve the two sequences in this optical CMD.} 
 \label{Bellini2} 
\end{figure} 
\end{centering}

An AS is considered successfully recovered if the difference between its input and output positions is less than 1\,pixel, the difference in magnitudes is within 0.75 (equivalent to $\sim$2.5log2) in all filters, and if it passes the same photometric quality and PM–based membership selection criteria applied to real stars. These criteria are consistent with the standard selections commonly adopted in AS tests in the literature \citep[e.g.][]{2008AJ....135.2055A,2008ApJ...678.1279B,2009ApJ...697..965B}, ensuring a reliable and homogeneous comparison between simulated and observed samples.

\begin{centering} 
\begin{figure}
 \includegraphics[width=\columnwidth]{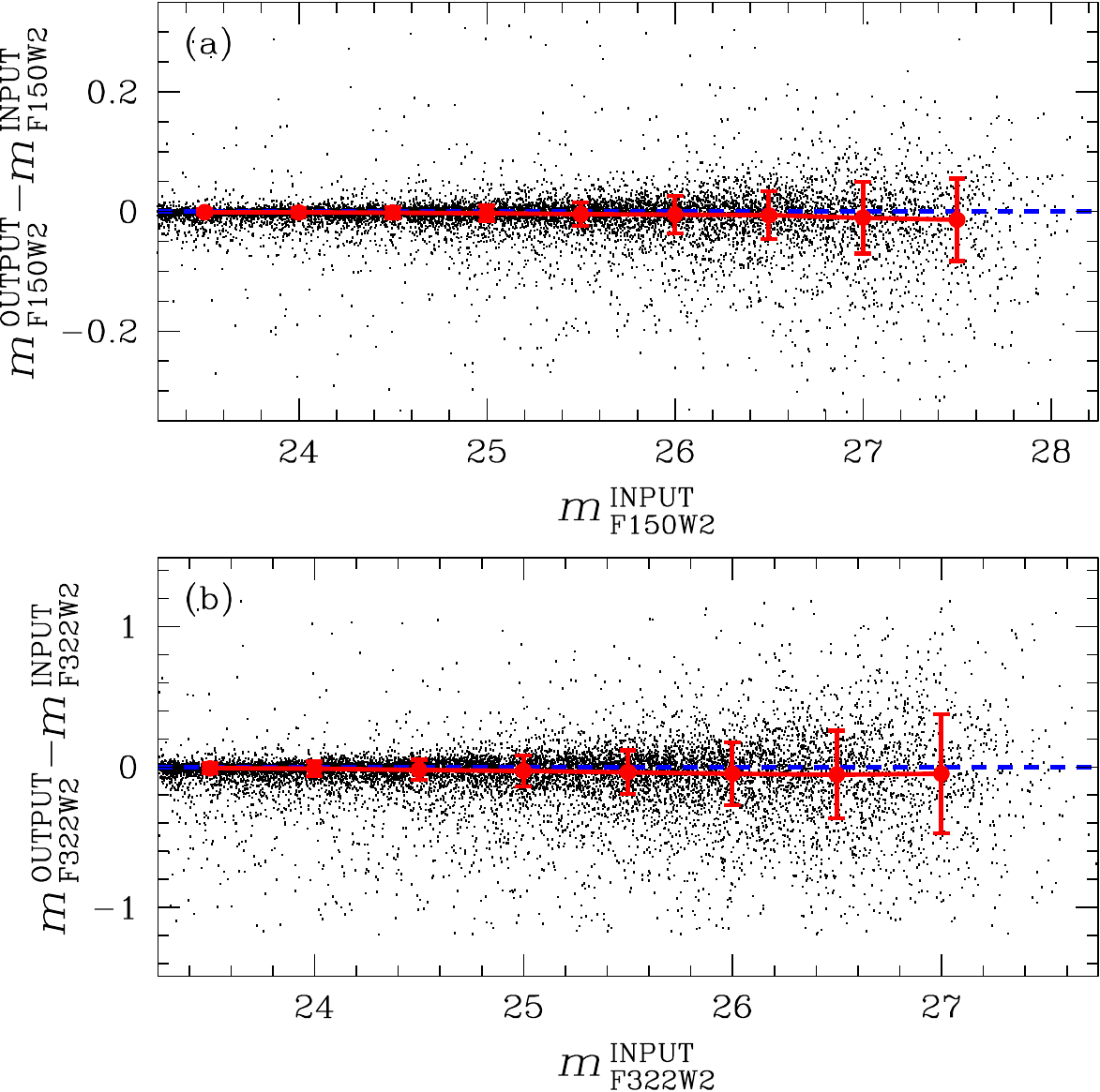}
 \caption{Photometric uncertainties estimated from ASs. Panel (a) shows the results for the F150W2 filters, while panel (b) refers to the F322W2 filter. In each panel, the black points represent the difference between the recovered and input magnitudes of the ASs, plotted as a function of input magnitude. The red points mark the median values in 0.5-mag bins, and the error bars indicate the corresponding dispersions ($\sigma$) obtained from the 2.5$\sigma$-clipped distributions. The dashed blue line marks zero.}
 \label{err_jwst} 
\end{figure} 
\end{centering}

\begin{centering} 
\begin{figure}
 \includegraphics[width=\columnwidth]{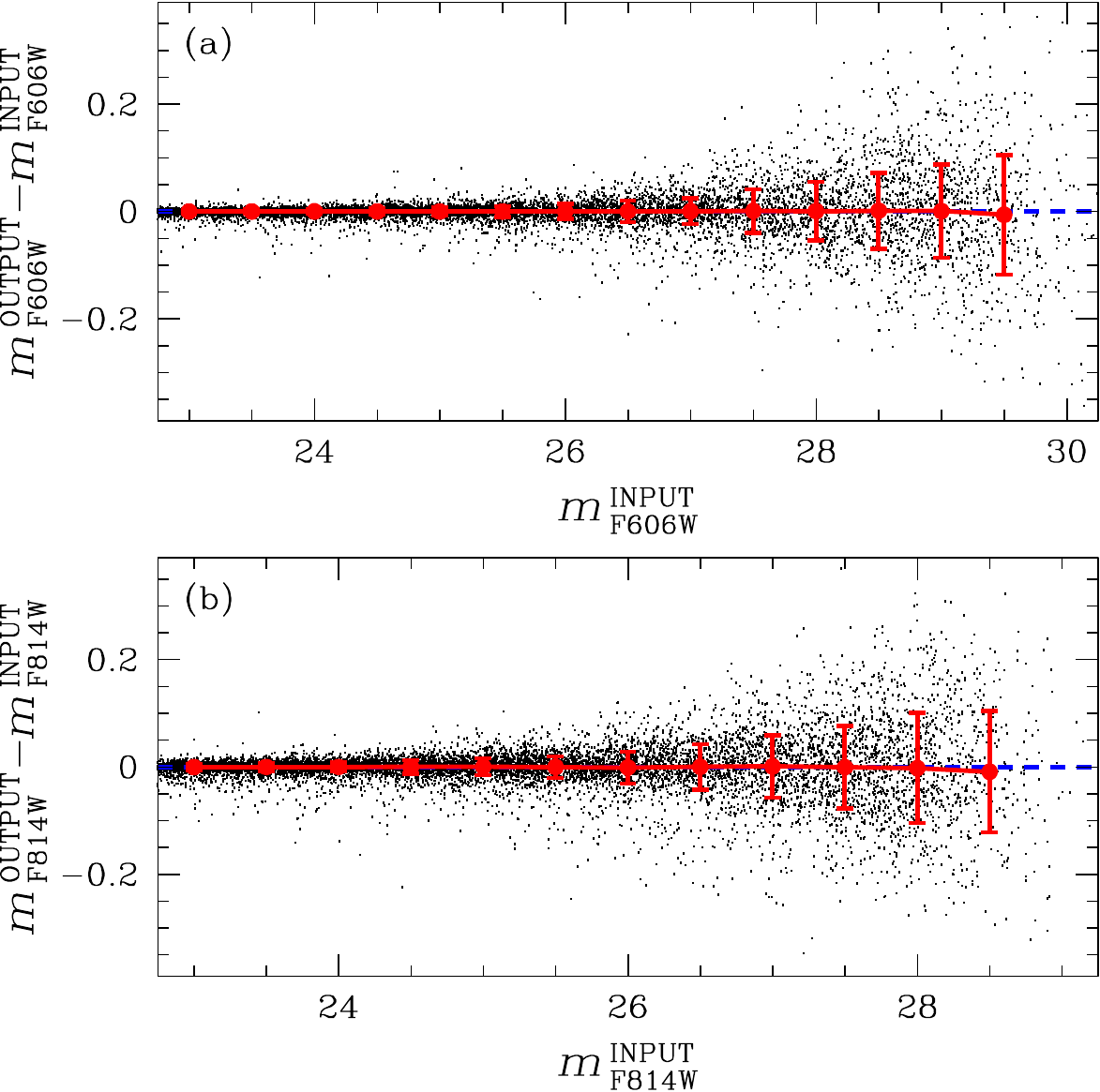}
 \caption{Same as Fig.\,\ref{err_jwst} but for the F606W (a) and F814W (b) filters.}
 \label{err_hst} 
\end{figure} 
\end{centering}

We used ASs to estimate the photometric uncertainties as a function of the magnitude in each of the four filters. The recovered ASs were grouped into 0.5-magnitude bins, and within each bin, we computed the 2.5$\sigma$-clipped median of the differences between the injected and recovered magnitudes. The final $\sigma$ from this clipping procedure was adopted as our estimate of the photometric error. 
The results of this procedure are shown in Fig.\,\ref{err_jwst} and \ref{err_hst}, which illustrate the photometric uncertainties derived from ASs for each filter. The uncertainties were estimated down to the faintest magnitudes where ASs are still successfully recovered, and interpolated for fainter magnitudes.

\section{The white dwarf cooling sequence of $\omega$\,Cen}\label{Section4}

As discussed in Section\,\ref{Section1}, the study by \citet{2013ApJ...769L..32B} revealed that the upper portion of the WD CS in $\omega$\,Cen splits into two distinct sequences. These were interpreted as a blue WD CS, mainly composed of standard-mass CO core WDs with masses around $0.53$–$0.55\,M_{\odot}$, and a red WD CS, consisting of lower-mass WDs, possibly a combination of objects with CO and He cores. This discovery was made possible through the use of ultraviolet photometry, obtained with the ultraviolet and visible (UVIS) channel of the Wide Field Camera 3 (WFC3) onboard HST. In particular, their analysis was based on observations in the F275W, F336W, and F438W filters, covering a central field in the cluster.

This result is shown in Fig.\,\ref{Bellini1}, which displays the $m_{\rm F275W}$ versus $m_{\rm F275W} - m_{\rm F438W}$ CMD for the central region of $\omega$\,Cen, using the publicly available catalogue presented and released by \citet{2014ApJ...797..115B}. From this dataset, we selected a sample of well-measured stars by applying photometric quality selections based on the \texttt{RADXS} parameter, retaining only sources with $|\texttt{RADXS}| < 0.05$ in all three bands. 

In this dense central field, the bright and hot end of the sequence is well populated by WDs with $T_{\rm eff}$ exceeding $\sim 15{,}500$\,K. The WD CS exhibits a distinct bifurcation in its upper portion, consistent with the results by \citet{2013ApJ...769L..32B}. The split is visible starting from $m_{\rm F275W} \sim 19.5$ and continues down to $m_{\rm F275W} \sim 23.5$. 

The two branches are separated by $\approx$0.1–0.2 magnitudes in colour along this interval. To quantify the populations along the two branches, we manually defined a fiducial line separating them, and colour-coded the stars along the two sequences in blue and red. Based on this selection, we identified 508 WDs in the red sequence and 292 in the blue one, yielding a relative fraction in agreement with the values reported by \citet{2013ApJ...769L..32B}.

It is important to emphasise that the separation between the two WD sequences is visible only when using ultraviolet filters in the dataset employed by \citet{2013ApJ...769L..32B,2014ApJ...797..115B}. The precision of the optical photometry in this central field, for these datasets, is not sufficient to disentangle the two populations. To demonstrate this, Fig.\,\ref{Bellini2} shows the $m_{\rm F606W}$ versus $m_{\rm F606W} - m_{\rm F814W}$ CMD of the stars displayed in Fig.\,\ref{Bellini1}, using the same colour-coding for the two WD sequences. This photometry, also from the \citet{2014ApJ...797..115B} catalogue, clearly shows that the two groups of WDs -- which appeared separated in the ultraviolet CMD -- are completely blended in the optical CMD. No trace of the previously observed split can be discerned in this diagram, suggesting that ultraviolet data are essential to reveal this feature. This is mainly due to the contamination from bright neighbouring stars, particularly RGB and MS stars, the latter being significantly brighter than even the brightest WDs in the optical. Such contamination, which severely affects the optical bands, is largely negligible in the ultraviolet, where crowding effects are substantially reduced.

More recently, \citet{2024A&A...691A..96S} carried out a detailed analysis of the WD CS of $\omega$\,Cen in a much more external region of the cluster. This study, which took advantage of an extensive HST dataset comprising 132 orbits, was able to identify the cluster's faintest members, down to the luminosity-function drop at the bottom of the WD CS. In this paper, we revisit this external field, combining the HST dataset from \citet{2024A&A...691A..96S} with new JWST NIR observations. The dataset used in \citet{2024A&A...691A..96S} included just optical imaging in the F606W and F814W filters, with no ultraviolet photometry. In these bands, the WD CS appears as a single, continuous sequence with no obvious evidence of a bifurcation, as shown in the $m_{\rm F606W}$ versus $m_{\rm F606W} - m_{\rm F814W}$ CMD presented in panels\,(d), (f) and (g) of Fig.\,\ref{pm}.

Note that in the external field studied here, essentially all cluster WDs are fainter than $m_{\rm F606W} \simeq 25$ (see panels\,(d), (f), and (g) of Fig.\,\ref{pm}). In contrast, the split in the WD CS reported by \citet{2013ApJ...769L..32B} and reproduced in Fig.\,\ref{Bellini1} involves WDs brighter than this limit (see Fig.\,\ref{Bellini2}). Although incompleteness reduces the number of faint WDs in Fig.\,\ref{Bellini2}, the central field hosts a much larger population of bright WDs ($m_{\rm F606W} < 25$) compared to Fig.\,\ref{pm}, simply because it is $\sim 1000$ times more populous than the external field. Despite the significantly lower crowding and longer exposure times in the dataset from \citet{2024A&A...691A..96S}, these improvements alone were not sufficient to reveal the split in the same upper portion of the WD CSc, as seen in the more crowded central-field observations of \citet{2013ApJ...769L..32B,2014ApJ...797..115B}, simply because the insufficient statistic of WDs brigher than $m_{\rm F606W} \simeq 25$.

The JWST data first reported in \citetalias{2025A&A...701A.169S}, obtained for the same field analysed in \citet{2024A&A...691A..96S} now offer a valuable opportunity to revisit this issue, and potentially extend the investigation to fainter magnitudes along the WD CS. Thanks to these NIR observations, we can explore alternative photometric colour combinations to assess whether the split can also be detected in this external field and at significantly lower $T_{\rm eff}$.

\begin{centering} 
\begin{figure}
 \includegraphics[width=\columnwidth]{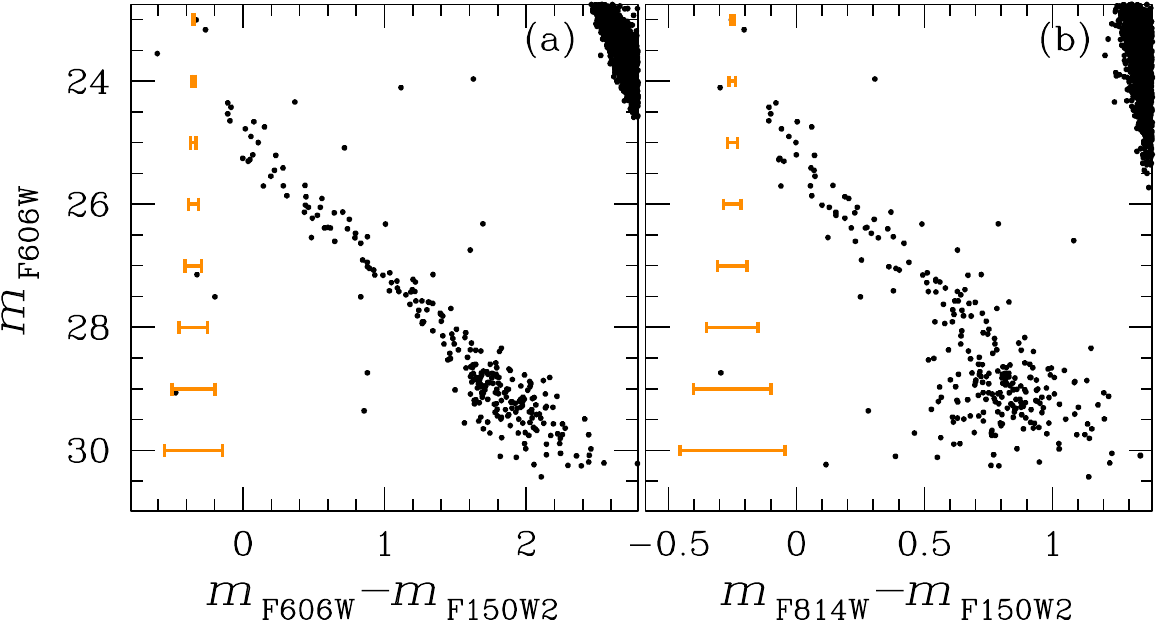}
 \caption{$m_{\rm F606W}$ versus $m_{\rm F606W} - m_{\rm F150W2}$ (panel a) and $m_{\rm F606W}$ versus $m_{\rm F814W} - m_{\rm F150W2}$ (panel b) CMDs of our selected sample of sources, focusing on the WD CS. The orange error bars on the left side of each panel indicate the photometric uncertainties in colour, derived from ASs, as a function of the $m_{\rm F606W}$ magnitude.}
 \label{cmd} 
\end{figure} 
\end{centering}

Figure\,\ref{cmd} shows the CMDs of sources that pass both the photometric quality and PM selection presented in Section\,\ref{Section2}. Panel\,(a) displays the $m_{\rm F606W}$ versus $m_{\rm F606W} - m_{\rm F150W2}$ CMD, while panel\,(b) shows the $m_{\rm F606W}$ versus $m_{\rm F814W} - m_{\rm F150W2}$ CMD. In both diagrams, the WD CS exhibits an unusual broadening in the magnitude range $25<m_{\rm F606W}<26.7$. This broadening may potentially trace the presence of the same double sequence originally identified in the ultraviolet within the central field analysed by \citet{2013ApJ...769L..32B}.

\begin{centering} 
\begin{figure}
 \includegraphics[width=\columnwidth]{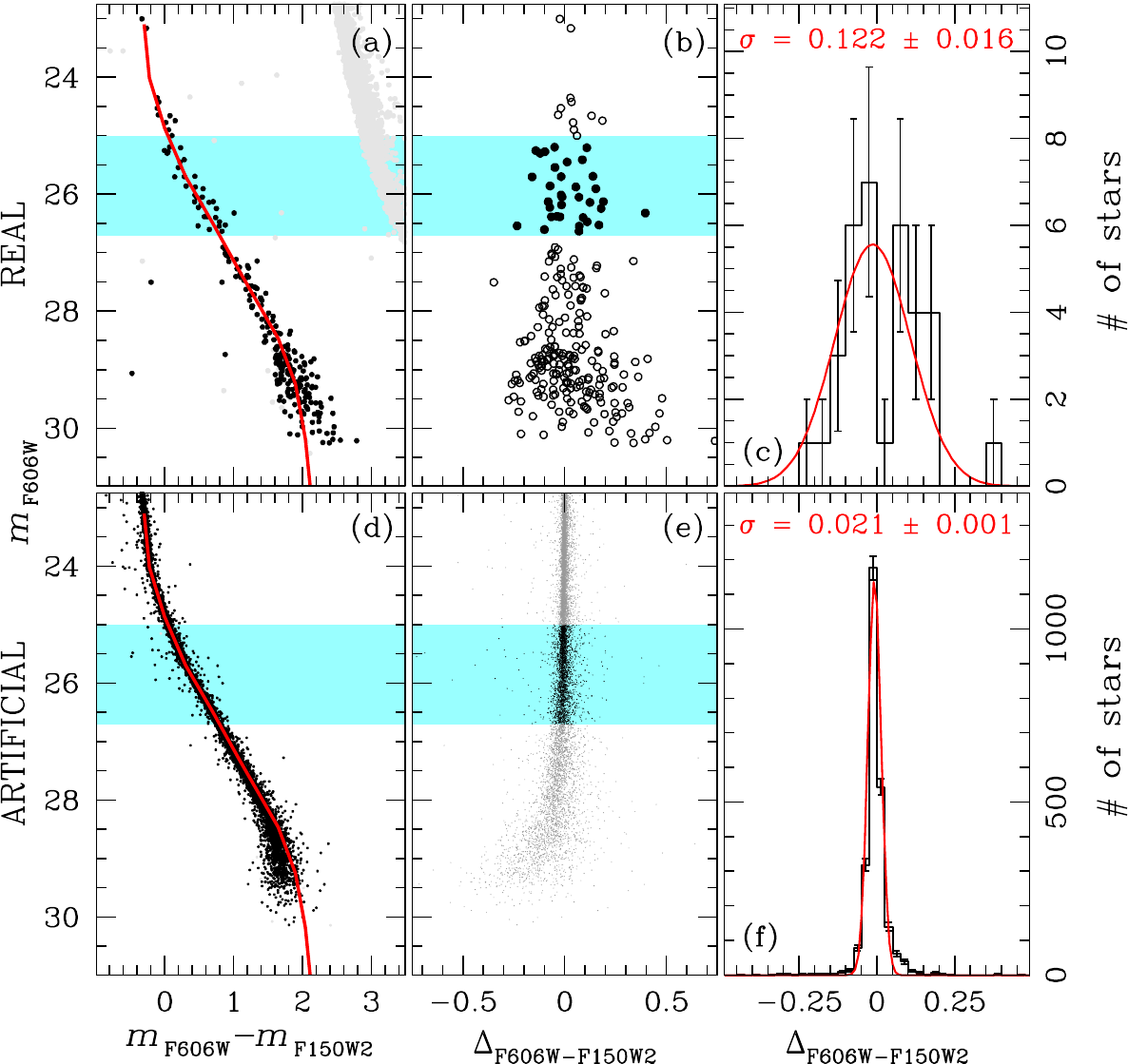}
 \caption{Comparison between the observed and AS distributions. The top panels show the analysis for real stars, while the bottom panels show the corresponding analysis for ASs. Black points indicate the sources located between the two green fiducial lines in panel\,(f) of Fig.\,\ref{pm}, whereas grey points mark the remaining stars. (a)-(d) $m_{\rm F606W}$ versus $m_{\rm F606W} - m_{\rm F150W2}$ CMDs, where the red line represents the fiducial sequence used to inject ASs. (b)-(e) Verticalised CMDs based on the fiducial sequence. $\Delta_{\rm F606W-F150W2}$ represents the verticalized colour. (c)-(f) Histograms of the colour distribution for sources within the cyan-shaded regions of the verticalised CMDs. The red curves show Gaussian fits, with the dispersion ($\sigma$) computed from the 68.27th percentile of the distributions. The corresponding $\sigma$ values are indicated in each panel.} 
 \label{sig} 
\end{figure} 
\end{centering}

We used ASs to demonstrate that the observed dispersion along the WD CS is significantly larger than the effect of our estimated uncertainties; the analysis is presented in Fig.\,\ref{sig}. The top panels refer to real sources, while the bottom panels correspond to ASs. For this analysis, we considered only the sources located between the two green fiducial lines shown in panel\,(f) of Fig.\,\ref{pm}, for both the real stars and the ASs. These sources are represented in black in Fig.\,\ref{sig}, while the other stars are shown in grey.

Panels\,(a) and (d) show the $m_{\rm F606W}$ versus $m_{\rm F606W} - m_{\rm F150W2}$ CMDs for our selected sample of real stars and the corresponding diagram for ASs. The red fiducial line represents the reference sequence used to inject ASs. We used this fiducial to verticalize the CMDs, shown in panels\,(b) and (e).

We restricted our quantitative analysis to the magnitude interval $25<m_{\rm F606W}<26.7$, where the broadening is most clearly visible. At brighter magnitudes, the limited number of stars in the external field prevents a robust statistical analysis, while at fainter magnitudes, the increasing photometric errors reduce the contrast. Panels\,(c) and (f) display the histograms of the verticalized CMDs ---for the stars within the magnitude interval $25<m_{\rm F606W}<26.7$, indicated by the region shaded in cyan--- where the Gaussian fits are based on the dispersion measured as the 68.27th percentile of the distributions. The corresponding dispersion values ($\sigma$) are labelled in each panel. The observed dispersion for real stars ($\sigma = 0.122 \pm 0.015$) is much larger than that estimated for ASs ($\sigma = 0.021 \pm 0.001$), and the histogram of real stars suggests the presence of two groups of WDs separated by a gap. This result points to an intrinsic broadening of this portion of the WD CS in $\omega$\,Cen, which cannot be solely attributed to photometric uncertainties.

\begin{centering} 
\begin{figure}
 \includegraphics[width=\columnwidth]{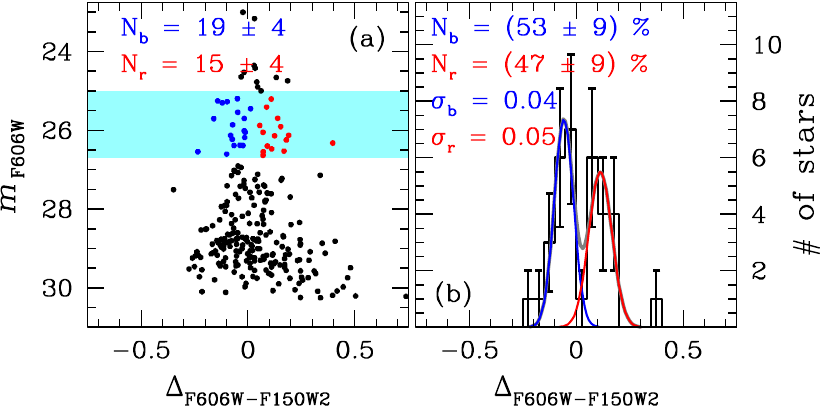}
 \caption{Estimate of the number of WDs in each population. (a) Verticalized $m_{\rm F606W}$ versus $\Delta_{\rm F606W-F150W2}$ CMD. The magnitude interval $25 < m_{\rm F606W} < 26.7$ adopted is highlighted in cyan. The two WD groups are shown in blue (bWD) and red (rWD), with their respective numbers of stars indicated. The uncertainties correspond to Poisson errors. (b) Histogram of the $\Delta_{\rm F606W-F150W2}$ distribution for stars within the cyan-shaded region of panel\,(a). The best-fit double-Gaussian model is plotted in grey, with the individual components in blue and red. The fractions of stars in each component are derived from the Gaussian areas, and the corresponding $\sigma$ values are reported.} 
 \label{2G} 
\end{figure} 
\end{centering}

In Fig.\,\ref{2G} we attempt to separate the two WD populations hinted at in the previous figure, and to estimate the number of WDs in each group. The figure shows the same verticalised CMD and histogram shown in panels\,(b) and (c) of Fig.\,\ref{sig}, respectively. 

From the verticalised CMD in panel\,(a), we selected WDs within the magnitude interval $25 < m_{\rm F606W} < 26.7$, highlighted by the cyan-shaded region, and divided them into two groups: stars with $\Delta_{\rm F606W-F150W2} \gtrsim 0.05$ are classified as red WDs (rWDs), while those with $\Delta_{\rm F606W-F150W2} \lesssim 0.05$ are classified as blue WDs (bWDs). Based on this selection, we identified 19 bWDs and 15 rWDs.

As an alternative approach, in panel\,(b), we fitted the histogram using a double-Gaussian model (grey curve), with the individual components shown in blue and red. From the area under each Gaussian component, we estimate that bWDs constitute approximately (53$\pm$9)\% of the sample, while rWDs account for the remaining (47$\pm$9)\%. The uncertainties are Poisson errors propagated through standard error propagation.

\begin{centering} 
\begin{figure}
 \includegraphics[width=\columnwidth]{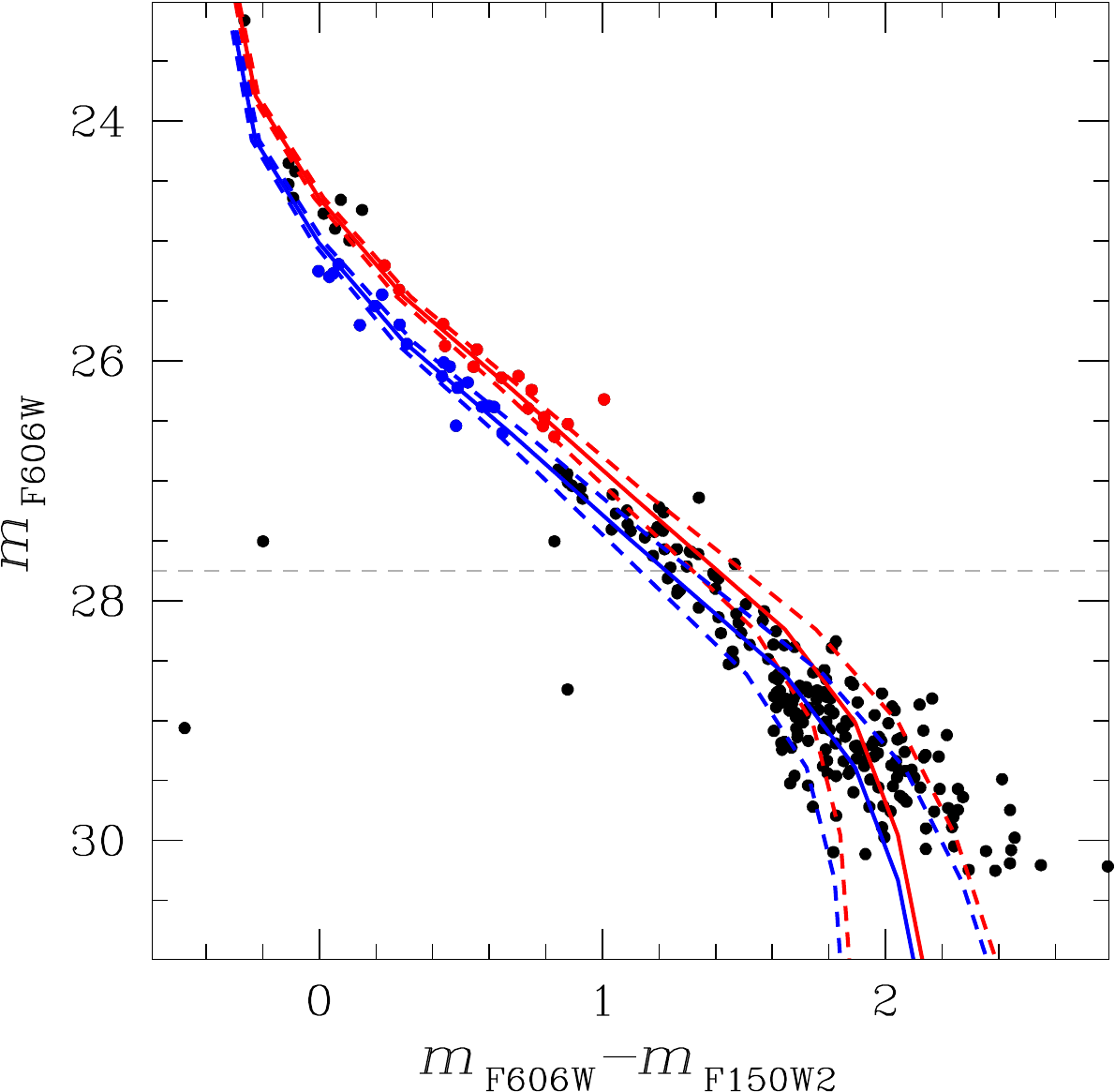}
  \caption{$m_{\rm F606W}$ versus $m_{\rm F606W} - m_{\rm F150W2}$ CMD for the selected sample of WDs, with stars colour-coded as in panel\,(a) of Fig.\,\ref{2G}. The red and blue solid lines represent fiducials for the two sequences. Dashed lines denote the $\pm1\sigma$ colour uncertainty (from ASs tests) around each fiducial. The grey dashed horizontal line marks the magnitude below which the two sequences can no longer be separated, as the colour difference between the fiducials becomes comparable to the photometric errors.}
 \label{merg} 
\end{figure} 
\end{centering}

Figure\,\ref{merg} shows a test to verify whether the two sequences become indistinguishable at fainter magnitudes due to increasing photometric errors, or whether the red sequence actually disappears because low-mass WDs are no longer present beyond a certain magnitude (as expected if the red sequence originates from low-mass WDs descending from the extreme HB, which was absent at earlier epochs in the cluster’s history).

The figure displays the $m_{\rm F606W}$ versus $m_{\rm F606W} - m_{\rm F150W2}$ CMD of our selected WD sample, with bWD and rWD stars colour-coded in blue and red, respectively, as in panel\,(a) of Fig.\,\ref{2G}. We overplotted two fiducial lines, shown in blue and red, which trace the two sequences. These fiducials were derived from the one shown in panel\,(a) of Fig.\,\ref{sig}, with small offsets applied to match the loci of the two sequences. The two fiducials differ in colour by $\sim 0.15$ mag in the region where the sequences are most clearly separated ($25 < m_{\rm F606W} < 26.7$).

Dashed lines around each fiducial denote the $\pm1\sigma$ photometric uncertainties in colour, as a function of $m_{\rm F606W}$, estimated from the ASs tests. The colour separation between the fiducials remains above the 1$\sigma$ uncertainty down to $m_{\rm F606W} \sim 27.75$, where the distance between the sequences becomes comparable to the photometric errors. Below this magnitude, the two sequences begin to merge and become fully indistinguishable. This confirms that, if still present, the two sequences can no longer be disentangled below $m_{\rm F606W} \sim 27.75$ due to photometric limitations. In other words, the current data do not allow us to determine whether the WD CS split continues at fainter magnitudes or fully merges, as we simply lack the colour resolution to explore below $m_{\rm F606W} \sim 27.75$.

Figure\,\ref{cmd2} in Appendix\,\ref{Appendix:A} displays a collection of CMDs of the WD CS in $\omega$\,Cen, constructed using various combinations of the HST and JWST filters available for this study. In each diagram, the two WD populations identified in panel\,(a) of Fig.\,\ref{2G} are highlighted in red and blue. The two groups trace recognisably distinct sequences across most of the CMDs, including the $m_{\rm F606W}$ versus $m_{\rm F606W} - m_{\rm F814W}$ diagram. However, it is important to emphasise that this separation would not have been readily apparent without the prior identification of the two populations based on their $m_{\rm F606W} - m_{\rm F150W2}$ colour, as presented in Fig.\,\ref{2G}. In other words, while the two sequences appear in multiple filter combinations, their distinction only becomes evident once the classification has been carried out using a combination of optical+NIR colour baseline, which enhances the separation between the populations. This enhanced visibility primarily arises because, at these magnitudes, the photometric errors in this particular colour combination ($m_{\rm F606W} - m_{\rm F150W2}$) are smaller than in others.

As a final note, we point out that the magnitude interval explored in this work ($25 < m_{\rm F606W} < 26.7$) differs from that analysed by \citet{2013ApJ...769L..32B}, who focused on a brighter portion of the WD CS ($22 < m_{\rm F606W} < 25$, see Fig\,\ref{Bellini2}). Our study extends to fainter luminosities (and older cooling ages), where the split manifests as a broadening in the optical+NIR CMDs and where photometric quality and the low crowding are still sufficient to investigate the presence of multiple WD populations.

\subsection{Theoretical interpretation of the WD sequence broadening}\label{Section4.1}

The analysis presented above demonstrates that the observed broadening in the WD CS of $\omega$\,Cen, particularly between $25 < m_{\rm F606W} < 26.7$ (see Fig.\,\ref{cmd}), is statistically significant and cannot be attributed solely to photometric uncertainties. This feature, detected here in CMDs combining HST optical and JWST NIR photometry, suggests it may be the same phenomenon observed in the even brighter portion of the WD CS using HST ultraviolet filters in a central field of $\omega$\,Cen by \citet{2013ApJ...769L..32B}, namely a split caused by two distinct WD populations originating from the cluster's known chemically distinct stellar populations. To test this hypothesis, we compare our new optical and NIR data with theoretical WD CSs representing these two populations.

Following \citet{2013ApJ...769L..32B}, we associate the two sequences with canonical CO-core WDs descending from the cluster's He-normal population and lower-mass WDs (a mix of He-core and CO-core) originating from the He-rich population. We employ the BASTI cooling models \citep{2022MNRAS.509.5197S} and Montreal hydrogen model atmospheres  \citep{1995PASP..107.1047B,2011ApJ...730..128T,2018ApJ...863..184B} for two representative populations: a sequence of $0.54$\,M$_{\odot}$ CO-core WDs and a sequence of $0.46$\,M$_{\odot}$ He-core WDs (dominant component from He-rich progenitors, \citealt{2013ApJ...769L..32B}). We adopt the cluster parameters of \citet{2024A&A...691A..96S}, specifically a distance modulus $(m - M)_0 = 13.67$ and a mean reddening $E(B-V) = 0.12$. Filter-specific extinctions were calculated using $A_V = 3.1 \times E(B-V)$ and the extinction ratios for JWST/NIRCam from \cite{2019ApJ...877..116W} ($A_{\rm F150W2} = 0.15 A_V$, $A_{\rm F322W2} = 0.04 A_V$) and appropriate coefficients for HST/ACS WFC F606W and F814W \citep{2005MNRAS.357.1038B}.

\begin{centering} 
\begin{figure*}
 \includegraphics[width=\textwidth]{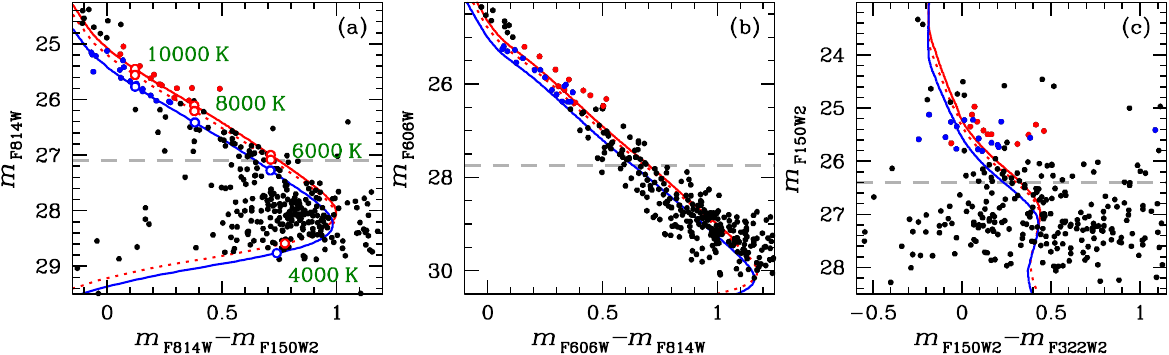}
 \caption{Comparison between the observed WD CS and theoretical cooling tracks. (a) ${m}_{\rm F814W}$ versus ${m}_{\rm F814W} - {m}_{\rm F150W2}$ CMD. (b) ${m}_{\rm F606W}$ versus ${m}_{\rm F606W} - {m}_{\rm F814W}$ CMD. (c) ${m}_{\rm F150W2}$ versus ${m}_{\rm F150W2} - {m}_{\rm F322W2}$ CMD. In all panels, red and blue points represent the bWD and rWD populations identified in panel\,(a) of Fig.\,\ref{2G}, respectively. The red solid and dotted lines show the cooling tracks of $0.46\,M_{\odot}$ He-core and CO-core WDs, respectively. The blue solid line shows the cooling track of a $0.54\,M_{\odot}$ CO-core WD. In panel\,(a), selected effective temperatures are indicated along the tracks. In each panel, the grey dashed horizontal line marks the magnitude below which the two sequences can no longer be separated, as their colour difference becomes comparable to the photometric errors.}
 \label{ct} 
\end{figure*} 
\end{centering}

As shown in Fig.\,\ref{ct}, the models provide a good match to the observed WD CS features. The predicted separation between the $0.54\,M_{\odot}$ and $0.46\,M_{\odot}$ sequences is approximately 0.1 magnitudes. This theoretically predicted separation quantitatively matches the observed broadening/split in our data across the relevant magnitude range, supporting the hypothesis. Note that this would imply the split to continue down to temperatures even cooler than $5{,}000$\,K. The comparison with single-mass cooling tracks is intended only to show the expected colour separation between CO-core and He-core WDs in the magnitude range relevant to our analysis. At fainter magnitudes, increasingly more massive WDs are predicted to populate the CS, and the single-mass approximation is no longer valid.

In addition, we note that, while the split in the CS of field WDs has been attributed mainly to different atmospheric compositions \citep[H- versus He-dominated;][]{2023MNRAS.523.3363B}, this explanation cannot account for the case of $\omega$\,Cen. As already shown by \citet{2013ApJ...769L..32B}, and confirmed by our results, the split observed here is best understood as arising from two populations of different WD masses associated with the cluster’s mPOPs.

The physical origin of this magnitude difference lies primarily in the different radii of the two WD populations. At a given effective temperature, the lower-mass ($0.46\,M_{\odot}$) WDs have significantly larger radii than the higher-mass ($0.54\,M_{\odot}$) WDs (15\% larger at $8{,}000$\,K). Since luminosity scales as $L = 4 \pi R^2 \sigma T_{\rm eff}^4$, the larger radius translates directly into a higher luminosity (32\% brighter at $8{,}000$\,K). The net effect in colour is that at a fixed luminosity, the He-core WD\,CS is about 0.1\,magnitudes redder than the canonical CO-core WD\,CS.

In panel\,(a) of Fig.\,\ref{ct}, selected effective temperatures are marked along the theoretical cooling tracks. The WDs identified as belonging to the two sequences span a temperature range $8{,}000 < T_{\rm eff} < 12{,}000$\,K in case of the rWDs, and $9{,}000 < T_{\rm eff} < 14{,}000$\,K in case of the bWDs, according to their respective models. While \citet{2013ApJ...769L..32B} were able to disentangle two WD groups down to $T_{\rm eff} \simeq 15{,}500$\,K, the optical-NIR combination used here allows us to trace the split between the two sequences down to $T_{\rm eff} \simeq 8{,}000$\,K, i.e., to stars evolved to the WD stage significantly earlier in time. In terms of WD ages, with \citet{2013ApJ...769L..32B} data, the two sequences can be disentangled until cooling ages of about 100-200\,Myr, while with our new data, the split can be traced until cooling ages of about 1\,Gyr. This implies that the earliest epoch of formation of this low-mass WD component is pushed back to more than 1\,Gyr ago.

Finally, in each panel, the grey dashed line indicates the magnitude below which the two sequences can no longer be clearly disentangled due to the increasing photometric uncertainties (as evaluated in the previous Section, see Fig.\,\ref{merg}). This threshold corresponds to $T_{\rm eff} \sim 6{,}000$\,K. In principle, the photometric precision of our catalogue would allow us to follow the split between the two populations down to this temperature; however, in practice, their separation becomes increasingly uncertain below $T_{\rm eff} \simeq 8{,}000$\,K, and we therefore restrict our analysis to the region where the split is most clearly visible.

\subsection{The radial gradient and kinematics}\label{Section4.2}

The two WD populations identified in $\omega$\,Cen are thought to be the descendants of the cluster’s two main MS components, as defined by \citet{2004ApJ...605L.125B}: the blue main sequence (bMS, helium-rich) and the red main sequence (rMS, helium-normal). In this framework, the bMS stars are expected to evolve into the rWDs, while the rMS stars give rise to the bWDs. 

Previous studies have shown that the bMS and rMS populations exhibit both a radial gradient and distinct kinematic properties. \citet{2007ApJ...654..915S} found that bMS stars are significantly more centrally concentrated than rMS stars. A study by \citet{2009A&A...507.1393B}, which combined HST and wide-field ground-based imaging to trace the cluster out to $\sim$20\,arcmin, showed that within the inner $\sim$2\,core radii, bMS stars slightly outnumber rMS stars. Beyond this region, the relative fraction of bMS stars steadily declines with increasing radius, flattening at $\sim$8\,arcmin and remaining constant at larger distances.

More recently, \citet{2024A&A...688A.180S} expanded on this work by combining multiple HST datasets covering a wide range of radial distances and filter combinations. Their analysis traced the radial distribution of the 15 distinct stellar populations previously identified in the core of $\omega$\,Cen by \citet{2017ApJ...844..164B}, effectively extending the study across nearly the entire cluster. In particular, \citet{2024A&A...688A.180S} confirmed earlier results, showing that bMS stars remain more centrally concentrated than their rMS counterparts throughout the cluster.

Comparing the radial and kinematic properties of the two WD populations with those of their MS progenitors offers a direct and compelling way to confirm their evolutionary link, while also shedding light on the origin and dynamical evolution of mPOPs in GCs.

\citet{2013ApJ...769L..32B} analysed the radial distribution of the two WD sequences in the central region of $\omega$\,Cen. They found that the fraction of rWDs -- associated with the more centrally concentrated bMS -- decreases with increasing radial distance. This radial trend was interpreted as the evolutionary imprint of the structural differences observed at the MS level, suggesting that at least part of the spatial segregation among the cluster's populations is preserved throughout the WD CS.

However, the spatial coverage of these earlier analyses was limited to the inner $\sim$150\,arcsec ($\sim$1 core radii), preventing any investigation of whether such a gradient persists at larger distances. The new field analysed in this study, where we also identify the two WD populations, is located much farther from the cluster centre -- at a radial distance comparable to the outer regions studied in \citet{2009A&A...507.1393B,2024A&A...688A.180S} for the MS populations -- and therefore offers a unique opportunity to extend the radial analysis of the WD populations into the outer cluster regions. This allows us to test whether the structural differences among $\omega$\,Cen's populations, observed at MS level, remain detectable also in their WD progeny over larger spatial scales.

\begin{centering} 
\begin{figure}
 \includegraphics[width=\columnwidth]{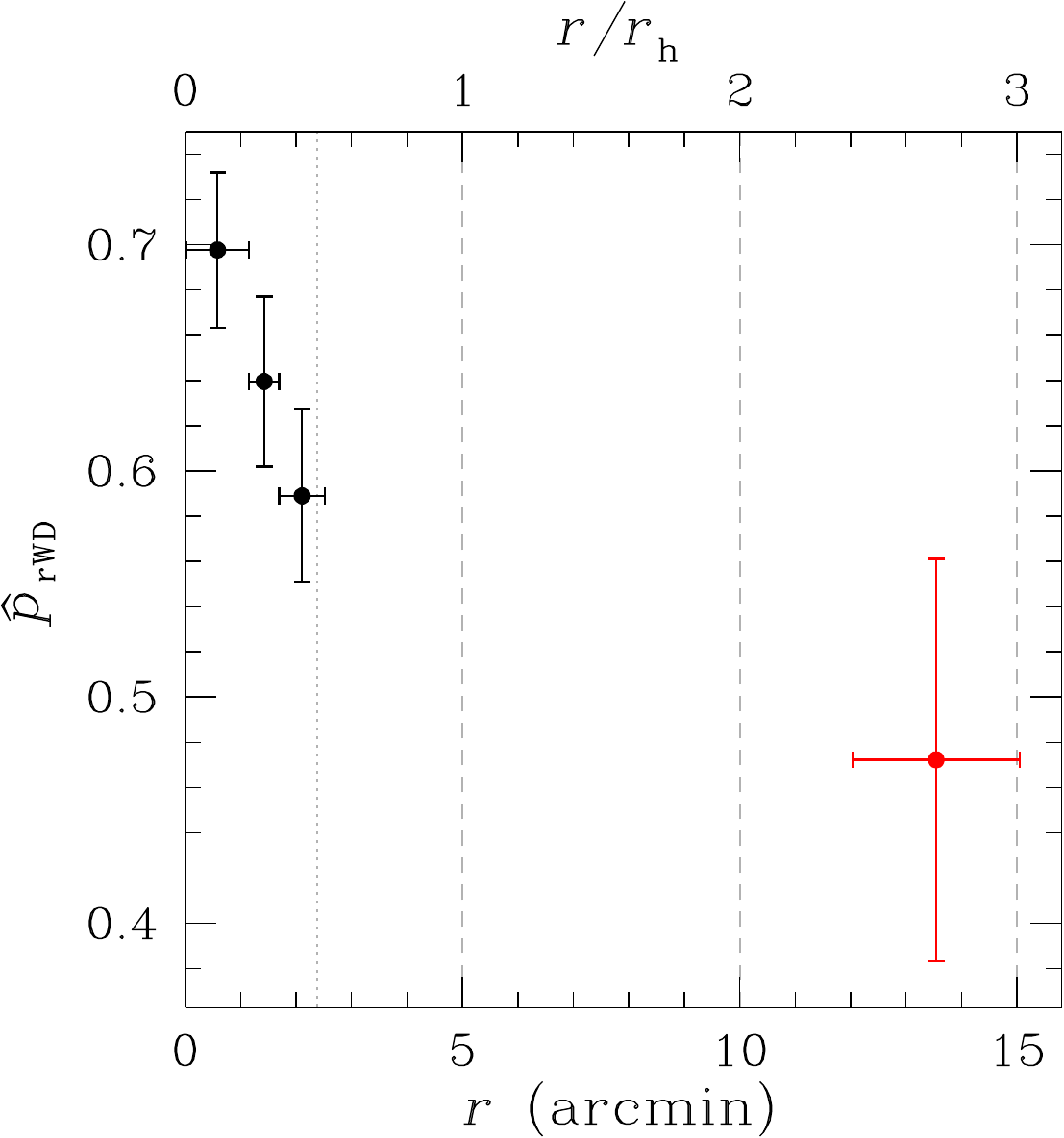}
 \caption{Radial distribution of the ratio of rWDs to the total WD population ($\hat{p}_{\rm rWD}$). Black points are from \citet{2013ApJ...769L..32B}, while the red point represents the results from this work. The dotted grey vertical line marks the core radius \citep[$r_{\rm c} = 2^\prime_{\cdot}37$][]{1996AJ....112.1487H, 2010arXiv1012.3224H}, and the dashed grey vertical lines indicate the half-light radius \citep[$r_{\rm h} = 5^\prime_{\cdot}00$][]{1996AJ....112.1487H, 2010arXiv1012.3224H} as well as 2\,$r_{\rm h}$ and 3\,$r_{\rm h}$.}
 \label{rad} 
\end{figure} 
\end{centering}

Figure\,\ref{rad} shows the radial distribution of $\hat{p}_{\rm rWD}$, the ratio of the number of rWDs to the total number of WD population in a given magnitude range. This distribution combines the results from \citet[][black points]{2013ApJ...769L..32B} with our own measurement (red point). As shown in the figure, our result qualitatively follows the trend identified by \citet{2013ApJ...769L..32B}, although in the three innermost bins the decline appears steeper, with $\hat{p}_{\rm rWD}$ decreasing as a function of distance from the cluster centre and reaching $\sim$0.47 in our outermost field.

The presence of a qualitatively similar radial gradient among the MS populations -- with bMS stars (the progenitors of rWDs) being more centrally concentrated than rMS stars -- provides further observational support for a direct connection between the double WD sequence and the well-known MS split observed in the optical \citep[see][]{2004ApJ...605L.125B}. This links the present-day properties of WDs to those of their progenitor populations.

We used PMs to investigate the dynamical properties of the two identified WD populations. For each population, we estimated the intrinsic velocity dispersions along the radial and tangential directions by maximising the following likelihood function:
\begin{align} 
    {\rm ln}L = - \frac{1}{2} \sum_{n} \Bigg[ & \frac{(v_{{\rm rad},n}-v_{\rm rad})^2}{\sigma_{\rm rad}^2+\epsilon_{{\rm rad},n}^2} + {\rm ln}(\sigma_{\rm rad}^2+\epsilon_{{\rm rad},n}^2) \nonumber \\ 
    &+ \frac{(v_{{\rm tan},n}-v_{\rm tan})^2}{\sigma_{\rm tan}^2+\epsilon_{{\rm tan},n}^2} + {\rm ln}(\sigma_{\rm tan}^2+\epsilon_{{\rm tan},n}^2) \Bigg],
    \label{eq:likelihood} 
\end{align}
where ($v_{{\rm rad},n}$, $v_{{\rm tan},n}$) and ($\epsilon_{{\rm rad},n}$, $\epsilon_{{\rm tan},n}$) are the radial and tangential PM components of the $n$-th star and their associated uncertainties, while ($v_{\rm rad}$, $v_{\rm tan}$) and ($\sigma_{\rm rad}$, $\sigma_{\rm tan}$) represent the mean motions and intrinsic velocity dispersions of each population. Uncertainties on ($v_{\rm rad}$, $v_{\rm tan}$) and ($\sigma_{\rm rad}$, $\sigma_{\rm tan}$) were estimated via bootstrap resampling with 1000 iterations.

\begin{centering} 
\begin{figure}
 \includegraphics[width=\columnwidth]{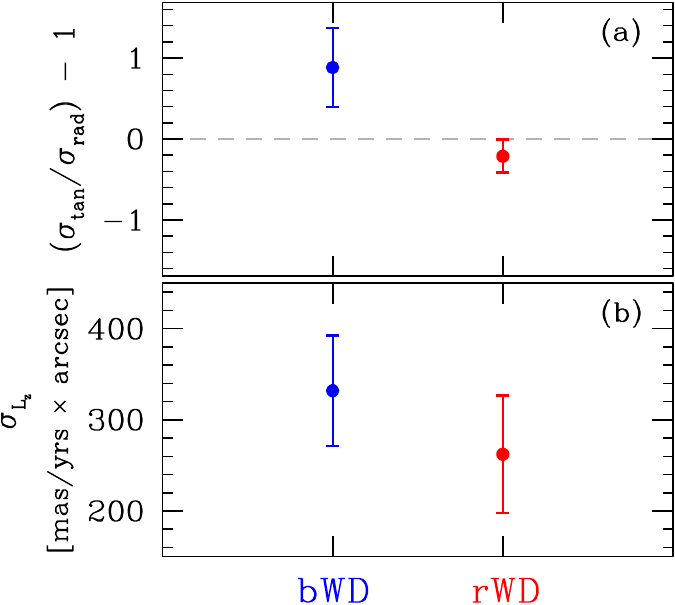}
 \caption{Dynamical properties of the two groups of WDs. (a) Velocity anisotropy for the two identified groups of WDs. The grey dashed line marks the 0. (b) Angular momentum of the two identified groups of WDs.}
 \label{dis} 
\end{figure} 
\end{centering}

Panel\,(a) of Fig.\,\ref{dis} shows the velocity anisotropy, defined as $(\sigma_{\rm tan}/\sigma_{\rm rad}) - 1$, for the two WD populations, with bWDs in blue and rWDs in red. Errors were propagated from the velocity-dispersion uncertainties. The figure reveals that while the bWD stars are characterised by a tangentially anisotropic velocity distribution, the rWD population has a slightly radially anisotropic velocity distribution.

To further investigate the dynamical differences between the two populations, we also measured the dispersion of the angular momentum $\sigma_{L_{z}}$ \citep[see e.g.][]{2025A&A...699A..44A,2025ApJ...986...80G}, where $L_{z,n} = v_{{\rm tan},n} \times r_n$ and $r_n$ is the projected distance of the $n$-th star from the cluster centre. As shown in \citet{2025A&A...699A..44A}, at a given distance from the cluster's centre, smaller values of  $\sigma_{L_{z}}$ indicate a more radially anisotropic velocity distribution. In our analysis, we derived the values of $\sigma_{L_{z}}$ for each population using a likelihood function analogous to Eq.\,\ref{eq:likelihood}. The results are shown in panel\,(b) of Fig.\,\ref{dis}.

A slightly smaller value of $\sigma_{L_{z}}$ is measured for the rWD stars, although the difference with respect to the bWDs is not statistically significant given the error bars. This is nevertheless consistent with the differences between the anisotropies in the velocity distribution suggested by panel\,(a) of Fig.\,\ref{dis}. As discussed in various studies \citep[see e.g.][]{2015ApJ...810L..13B,2021MNRAS.502.4290V,2025A&A...699A..44A}, the kinematic difference revealed by our analysis is consistent with that predicted by simulations of the dynamical evolution of multiple-population clusters in which the He-rich population is initially more centrally concentrated than the He-normal stars. This trend is also generally consistent with that found in the kinematic properties for multiple populations on the MS by \citet{2018ApJ...853...86B}. However, we emphasise that the relatively small number of WDs in the two samples introduces significant statistical uncertainties, and these results should therefore be interpreted with caution and further investigated in future studies.

\section{Conclusions}\label{Section6}

In this work, we have presented a detailed analysis of the WD CS in an outer field of the GC $\omega$\,Cen, combining optical data from HST with NIR imaging from JWST. Our study extends previous investigations of the signatures of mPOPs on the WD CS by probing both fainter magnitudes and larger radial distances.

Our main findings are as follows:

\begin{itemize}
\item We identified the presence of a significant broadening in the WD CS between $m_{\rm F606W} \sim 25$ and $26.7$ in CMDs combining HST and JWST filters. This broadening cannot be explained by photometric uncertainties alone, as demonstrated through ASs.

\item We interpret this broadening as the continuation to lower $T_{\rm eff}$ of the WD CS bifurcation first identified in the ultraviolet by \citet{2013ApJ...769L..32B}, and now detected for the first time in a low-crowding, outer field using optical+NIR photometry.

\item By verticalizing the CMD and modelling the colour distribution with a double-Gaussian fit, we identify two distinct WD populations, comprising (53$\pm$9)\% bWDs and (47$\pm$9)\% rWDs.

\item Theoretical comparisons with cooling models confirm that the observed separation is consistent with two distinct WD populations: canonical CO-core WDs with masses of $\sim$0.54\,$M_\odot$ (bWDs), and lower-mass He-core WDs \citep[with a possible small fraction of low-mass CO-core objects, see][]{2013ApJ...769L..32B} with masses around 0.46\,$M_\odot$ (rWDs), the latter likely descending from He-rich progenitor populations \citep[see discussion in][]{2013ApJ...769L..32B}.

\item The two sequences merge (because of the increasing photometric errors) at a magnitude corresponding to a cooling age of about 1\,Gyr. This pushes back the epoch of formation of the low-mass rWDs sequence to over 1\,Gyr ago, compared to the 100-200\,Myr derived from \citet{2013ApJ...769L..32B} results.

\item We find that the relative fraction of rWDs decreases with increasing radial distance from the cluster centre, extending the radial gradient first detected in the central regions by \citet{2013ApJ...769L..32B} to the outer field analysed here. This mirrors the behaviour observed among the progenitor MS populations, strengthening the evolutionary connection between the MS and WD populations in $\omega$\,Cen. In addition, our dynamical analysis indicates differences between the velocity distributions of bWDS and rWDs, where the velocity distribution of the bWDs is slightly tangential anisotropic, while that of the rWD population is slightly radially anisotropic. Although we emphasise that further investigation of the WDs' kinematics is necessary to reduce the statistical uncertainties, the trend hinted by our analysis is consistent with that previously found \citet{2018ApJ...853...86B} for MS stars and provides further dynamical evidence of the link between MS and WD stars in the two populations. 

\end{itemize}

This study demonstrates the power of combining HST and JWST observations to study the faintest WD populations in GCs. The detection of multiple WD sequences at large radial distances reinforces the scenario in which $\omega$\,Cen hosts chemically and structurally distinct stellar populations, whose signatures persist from the MS through the end stages of stellar evolution. Future JWST observations will be crucial to deepen this picture. In particular, our GO-5110 will provide a second epoch in 2026 with the filters F090W and F444W. These data will allow us to improve PM membership using the F090W filter, verify possible infrared excesses in F444W, and extend the analysis of the WD populations to even fainter magnitudes.

\begin{acknowledgements}
This research is based on observations made with the NASA/ESA Hubble Space Telescope and the NASA/ESA/CSA James Webb Space Telescope. The data were obtained from the Mikulski Archive for Space Telescopes (MAST) at the Space Telescope Science Institute (STScI), which is operated by the Association of Universities for Research in Astronomy, Inc., under NASA contracts NAS 5–26555 for HST and NAS 5-03127 for JWST. These observations are associated with programmes HST-GO-14118+14662 and JWST-GO-5110. The data described here may be obtained from \url{http://dx.doi.org/10.17909/7prx-4905}.
\end{acknowledgements}

\bibliographystyle{aa}
\bibliography{main.bib}

\begin{appendix}
\onecolumn
\section{Colour-magnitude diagrams of the white dwarf cooling sequence of $\omega$\,Cen}\label{Appendix:A}

\begin{centering} 
\begin{figure}[h!]
 \includegraphics[width=\columnwidth]{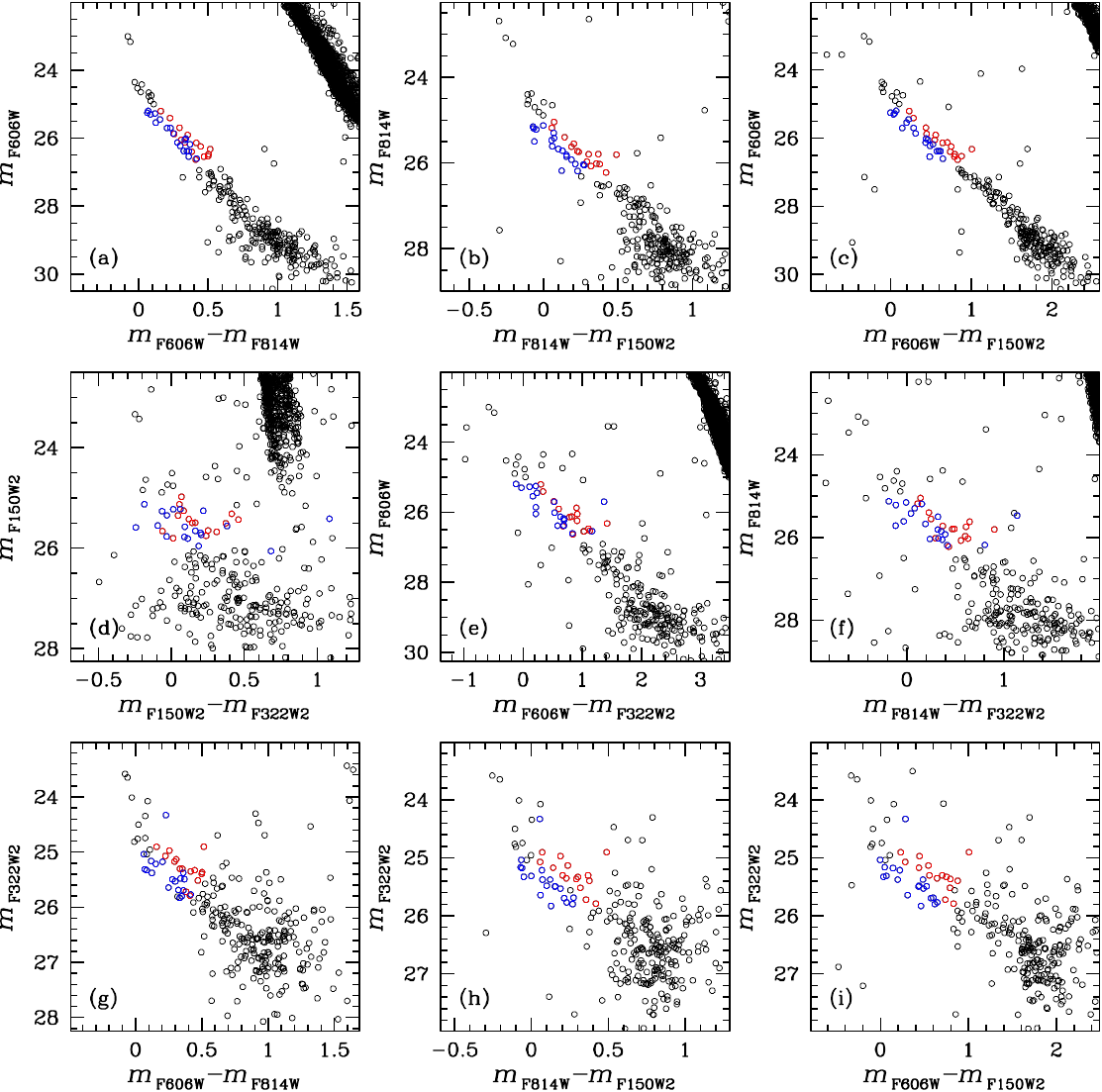}
 \caption{Nine CMDs of $\omega$\,Cen  WD CS, using the four HST and JWST filters employed in this study. The two identified sequences in this study are represented in blue and red.} 
 \label{cmd2} 
\end{figure} 
\end{centering}

In this Appendix, we provide a set of CMDs of the WD CS in $\omega$\,Cen, obtained by combining the four HST and JWST filters used in this work. These diagrams are shown in Fig.\,\ref{cmd2}, where the two identified WD sequences are highlighted in blue (bWD) and red (rWD).

\end{appendix}

\end{document}